\documentclass{ws-procs9x6}
\usepackage{amsmath}
\usepackage{graphicx}
\usepackage{amsfonts}
\usepackage{amssymb}

\begin{document}

\title{electron-positron-photon plasma around a collapsing star}
\author{R. RUFFINI, L. VITAGLIANO AND S.-S. XUE}
\maketitle

\address{ICRA and
Physics Department, \\
University of Rome ``La Sapienza", \\
P.le A. Moro 5, \\
00185 Rome, Italy}

\abstracts
{We describe electron-positron pairs creation around an electrically charged
star core collapsing to an electromagnetic black hole (EMBH), as well as pairs annihilation into photons. We use the kinetic Vlasov equation formalism for the pairs and photons and show that a regime of plasma oscillations is established around the core. As a byproduct of our analysis we can provide an estimate for the thermalization time scale.}

\section{Dynamics of Dyadosphere}

Dyadosphere was first introduced in Ref.~\refcite{PRX98} as the region
surrounding an electromagnetic black hole (EMBH) in which the electromagnetic
field strength exceeds the critical value $\mathcal{E}_{\mathrm{c}}$ for
electron-positron pair creation via the mechanism \emph{\`{a} l\`{a}
}Heisenberg-Euler-Schwinger.\cite{HE35,S51} The relevance of the dyadosphere
around an EMBH, for the astrophysics of gamma-ray busts has been discussed in
Refs.~\refcite{PRX98}, \refcite{RBCFX01a}--\refcite{RBCFX01c} (the external
radius of dyadosphere will be denoted by $r_{\mathrm{ds}}$). In those papers
the pair production in dyadosphere has been described as an electrostatic
problem: instantaneously a massive body collapses to an EMBH whose charge is
large enough that the electric field strength $\mathcal{E}$ exceeds
$\mathcal{E}_{\mathrm{c}}$ and the Schwinger process is triggered in the
entire dyadosphere; moreover the pairs are produced at rest and remain at rest
during the whole history of their production; finally they instantaneously
thermalize to a plasma configuration (see Fig.~\ref{dyaon}). These ansatz,
formulated for the sake of simplicity, allow one to estimate the number
density of pairs produced as well as the energy density deposited on the pairs
in a straightforward manner. We relax the hypothesis that the large electric
field is instantaneously built up and take the following dynamical point of view:

\begin{figure}[th]
\epsfxsize=7cm \centerline{\epsfxsize=3.9in\epsfbox{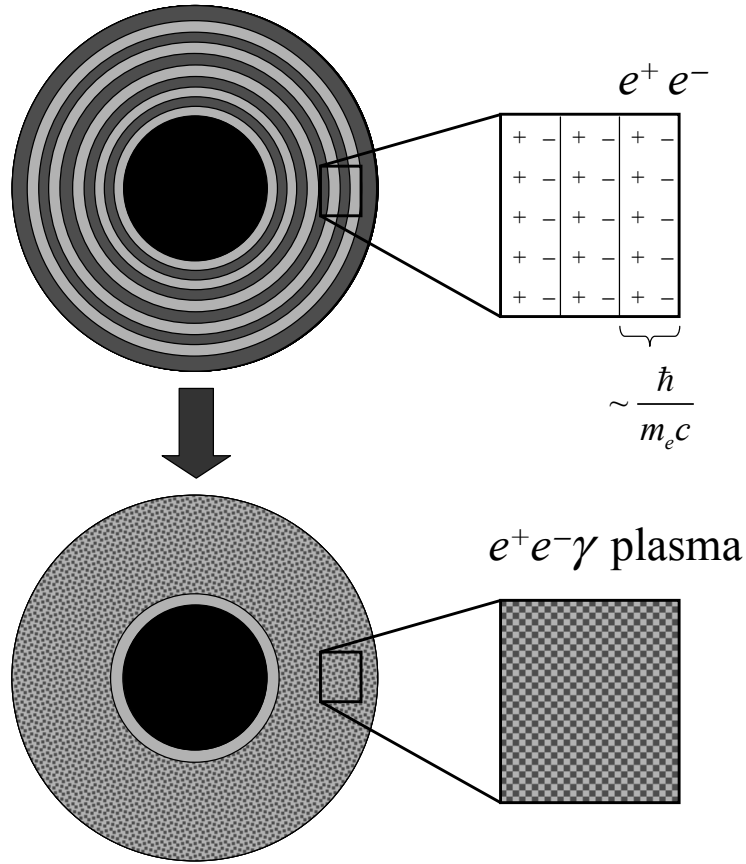}}\caption{Vacuum
polarization process of energy extraction from an EMBH. Pairs are created by
vacuum polarization in the dyadosphere and the system thermalizes to a neutral
plama configuration (see Ref.~[1] for details).}%
\label{dyaon}%
\end{figure}

\begin{enumerate}
\item  A spherically symmetric star core endowed with electric, say positive,
charge $Q$, collapses. We assume that the electromagnetic field strength
$\mathcal{E}$ on the surface of the core is amplified to $\mathcal{E}%
_{\mathrm{c}}$ during the collapse and the Schwinger process begins.

\item  The pairs produced by the vacuum polarization progressively screen the
electromagnetic field of the core, thus reducing its strength. Furthermore the
charges (electrons and positrons) are accelerated by the Lorentz force in the
electromagnetic field. Finally particles and antiparticles annihilate into photons.
\end{enumerate}

An enormous amount of pairs ($N\sim\tfrac{Q}{e}\tfrac{r_{\mathrm{ds}}}%
{\lambda_{\mathrm{C}}}$, where $\lambda_{\mathrm{C}}$ is the Compton length of
the electron) is produced, as claimed in Refs.~\refcite{PRX98}, \refcite
{RBCFX01a}--\refcite{RBCFX01c}, if the core charge is not annihilated by the
charge of the accelerated electrons during the gravitational collapse (see
Ref.~\refcite{RVX03}). Therefore it is useful to study the dynamics of the
electron-positron-photon plasma in the electric field of the core in some
details. This will be the main object of the next section. As a byproduct of
the analysis we obtain an estimate for the time scale needed for the
thermalization of the system.

\begin{figure}[th]
\epsfxsize=10cm \centerline{\epsfxsize=3.9in\epsfbox{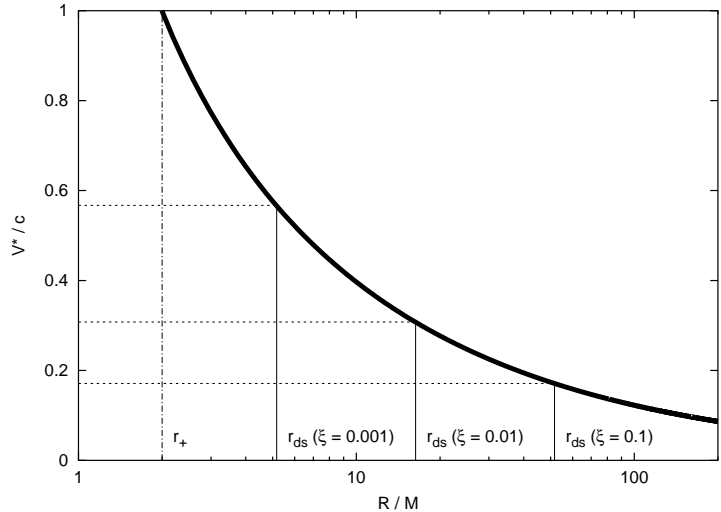}}%
\caption{Collapse velocity of a charged star of mass $M_{0}=M=20M_{\odot}$ as
measured by static observers as a function of the radial coordinate of the
star surface. As the charge is not too large ($\xi\lesssim0.1$) there is not
much difference between collapse velocities of stars with different charge.
Dyadosphere radii for different charge to mass ratios ($\xi=10^{-3}%
,10^{-2},10^{-1}$) are indicated in the plot together with the corresponding
velocity. }%
\label{Vstar}%
\end{figure}

In Ref.~\refcite{CRV02} it was suggested that the exact solution of
Einstein-Maxwell equations describing the gravitational collapse of a thin
charged shell can be used as a simplyfied analytical model for the
gravitational collapse of a charged core; it was also discussed in some
details the amplification of electromagnetic field strength on the surface of
the core. Here we briefly review some of the results of Ref.~\refcite{CRV02}.
The region of space-time external to the core is Reissner-Nordstr\"{o}m with
line element
\begin{equation}
ds^{2}=-fdt^{2}+f^{-1}dr^{2}+r^{2}d\Omega^{2}%
\end{equation}
in Schwarzschild like coordinate $\left(  t,r,\theta,\phi\right)  ,$ where
$d\Omega^{2}=d\theta^{2}+\cos^{2}\theta d\phi^{2}$, $f=f\left(  r\right)
=1-\tfrac{2M}{r}+\tfrac{Q^{2}}{r^{2}}$; $M$ is the total energy of the core as
measured at infinity and $Q$ is its total charge. Let us label with $R$ and
$T$ the radial and time-like coordinate of the shell, then the equation of
motion of the core is (cfr. Ref.~\refcite{CRV02})%

\begin{equation}
\tfrac{dR}{dT}=-\tfrac{f\left(  R\right)  }{H\left(  R\right)  }\sqrt
{H^{2}\left(  R\right)  -f\left(  R\right)  }\label{Motion}%
\end{equation}
where $H\left(  R\right)  =\tfrac{M}{M_{0}}-\tfrac{M_{0}^{2}+Q^{2}}{2M_{0}R}$;
$M_{0}$ is the rest mass of the shell. The analytical solution of
Eq.~(\ref{Motion}) was found in Ref.~\refcite{CRV02} in the form $T=T\left(
R\right)  $. According to a static observer $\mathcal{O}$ placed at the event
$x_{0}\equiv\left(  R,T\left(  R\right)  ,\theta_{0},\phi_{0}\right)  $ the
core collapses with speed given by
\begin{equation}
V^{\ast}\equiv-\tfrac{dR^{\ast}}{dT^{\ast}}=\sqrt{1-\tfrac{f\left(  R\right)
}{H^{2}\left(  R\right)  }}\leq1
\end{equation}
where $dR^{\ast}\equiv f^{-1/2}dR$ and $dT^{\ast}\equiv f^{1/2}dT$ are spatial
and temporal proper distances as measured by $\mathcal{O}$. In
Fig.~\ref{Vstar} we plot $V^{\ast}$ as a function of $R$ for a core with
$M=M_{0}=20M_{\odot}$\footnote{The condition $M_{0}=M$ corresponds to a shell
starting its collapse at rest at infinity.} and $\xi\equiv\tfrac{Q}{M}%
=10^{-3}$, $10^{-2}$, $10^{-1}$. Recall that dyadosphere radius is given
by\cite{PRX98}
\begin{equation}
r_{\mathrm{ds}}=\sqrt{\tfrac{eQ\hbar}{m_{e}^{2}c^{3}}},
\end{equation}
where $c$ is the speed of light; $e$ and $m_{e}$ are electron charge and mass
respectively. Then note that $V_{\mathrm{ds}}^{\ast}\equiv\left.  V^{\ast
}\right|  _{R=r_{\mathrm{ds}}}\simeq0.2c$ for $\xi=0.1$.

\section{Plasma oscillations and screening}

We now turn to the pair creation taking place during the gravitational
collapse. The gravitational fields of the core is considered classical; the
gravitational effects of the electron-positrons-photons plasma are neglected.

The most detailed framework for studying electromagnetic vacuum polarization
and particle-antiparticle scattering around an electromagnetic collapsing core
is quantum electrodynamics in the classical external electromagnetic field of
the core on the Reissner-Nordstr\"{o}m space-time around the core itself. Of
course a number of approximations is needed in order to make the problem be
tractable. Let us discuss such approximations.

\begin{enumerate}
\item [(Homogeneity)]First of all, the static Reissner-Nordstr\"{o}m
space-time region external to the collapsing core is naturally splitted in
space (hypersurfaces orthogonal to the static Killing field) and time. In the
local frames associated with static observers, the electromagnetic field of
the core is purely electric. Moreover, we will see that the length scale $L$
over which the electric field as well as the particle number densities vary,
is much larger than the length scale $l$ which is charachteristic of the
electron--positron motion. Thus we can divide dyadosphere into small regions
$D_{i}$
\begin{align}
D_{i} &  :r_{i}\leq r\leq r_{i+1}=r_{i}+\varepsilon;\label{D}\\
r_{+} &  \leq r_{i}\leq r_{\mathrm{ds}}\qquad\varepsilon\lesssim l\nonumber
\end{align}
such that for any $i$ the system formed by the electric field and the pairs
can be considered homogeneous in $D_{i}$.

\item[(Flat space-time)] For, in geometric units, the electron charge $e$ is
much larger than the electron mass $m_{e}$, the gravitational acceleration is
negligible with respect to electric acceleration for sufficiently large
electric field strengths (even much less than $\mathcal{E}_{c}$), therefore we
will neglect the curvature of space-time and use the local frame of a static
observer as a globally inertial frame of the Minkowski space-time.

\item[(Mean field)] The number of pairs is so high that a semiclassical
formalism and mean field approach can be used, in which the total
electromagnetic field (core electromagnetic field and screen field due to
pairs) is considered to be classical, while the electron-positron field is
quantized. It has been shown\cite{KESCM91}$^{-}$\cite{CEKMS93} that, if we
neglect scattering between particles, the semiclassical evolution of the
homogeneous system in a flat space-time is well described by a
Boltzmann-Vlasov-Maxwell system of partial differential equations, where the
electrons and positrons are described by a distribution function $f_{e}%
=f_{e}\left(  t,\mathbf{p}\right)  $ in the phase space, where $t$ is the
inertial time and $\mathbf{p}$ the $3-$momentum of electrons. Finally we use
the method presented in Ref.~\refcite{RVX02a} to simplify such a
Boltzmann-Vlasov-Maxwell system.
\end{enumerate}

Let us summarize results in Ref.~\refcite{RVX02a}: we obtained the following
system of ordinary differential equations which simultaneously describes the
creation and evolution of electron-positron pairs in a strong electric field
as well as the annihilation of pairs into photons:
\begin{equation}
\left\{
\begin{array}
[c]{l}%
\frac{d}{dt}n_{e}=S\left(  \mathcal{E}\right)  -2n_{e}^{2}\sigma_{1}\rho
_{e}^{-1}\left|  {\pi}_{e\parallel}\right|  +2n_{\gamma}^{2}\sigma_{2}\\
\frac{d}{dt}n_{\gamma}=4n_{e}^{2}\sigma_{1}\rho_{e}^{-1}\left|  {\pi
}_{e\parallel}\right|  -4n_{\gamma}^{2}\sigma_{2}\\
\frac{d}{dt}\rho_{e}=en_{e}\mathcal{E}\rho_{e}^{-1}\left|  {\pi}_{e\parallel
}\right|  +\tfrac{1}{2}\mathcal{E}j_{p}\left(  \mathcal{E}\right)  -2n_{e}%
\rho_{e}\sigma_{1}\rho_{e}^{-1}\left|  {\pi}_{e\parallel}\right|  +2n_{\gamma
}\rho_{\gamma}\sigma_{2}\\
\frac{d}{dt}\rho_{\gamma}=4n_{e}\rho_{e}\sigma_{1}\rho_{e}^{-1}\left|  {\pi
}_{e\parallel}\right|  -4n_{\gamma}\rho_{\gamma}\sigma_{2}\\
\frac{d}{dt}{\pi}_{e\parallel}=en_{e}\mathcal{E}-2n_{e}{\pi}_{e\parallel
}\sigma_{1}\rho_{e}^{-1}\left|  {\pi}_{e\parallel}\right| \\
\frac{d}{dt}\mathcal{E}=-2en_{e}\rho_{e}^{-1}\left|  {\pi}_{e\parallel
}\right|  -j_{p}\left(  \mathcal{E}\right)
\end{array}
\right.  , \label{sys}%
\end{equation}
where $n_{e}$ ($n_{\gamma}$) is the electron (photon) number-density,
$\rho_{e}$ ($\rho_{\gamma}$) is the electron (photon) energy-density, ${\pi
}_{e\parallel}$ is the density of electron radial momentum and $\mathcal{E}$
the electric field strength. Finally $S\left(  \mathcal{E}\right)  $ is the
Schwinger probability rate of pair creation, $j_{p}\left(  \mathcal{E}\right)
$ is the polarization current-density, $\sigma_{1,2}$ are total cross sections
for the processes $e^{+}e^{-}\rightleftarrows\gamma\gamma$ and the
corresponding terms describe probability rates of pair annihilation into
photons and vice versa. System (\ref{sys}) was numerically integrated in
Ref.~\refcite{RVX02a}.

\begin{figure}[th]
\epsfxsize=10cm \centerline{\epsfxsize=3.9in\epsfbox{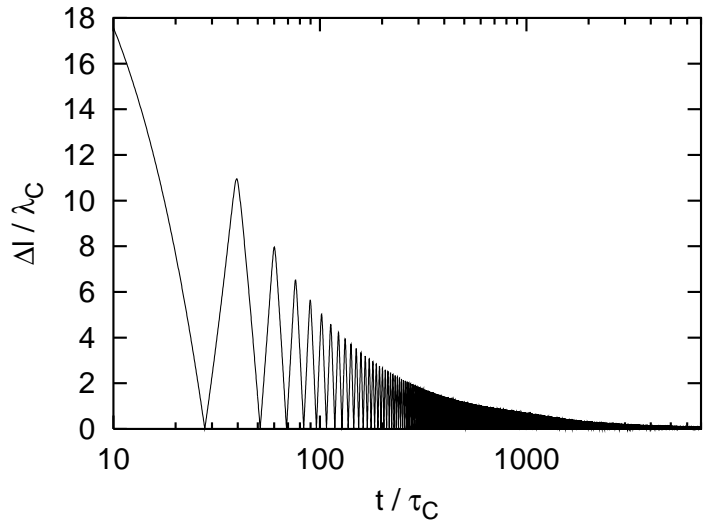}}%
\caption{Electrons elongation as function of time in the case $r=\tfrac{1}{3}
r_{\mathrm{ds}}$. The oscillations are damped in a time of the order of
$10^{3}\tau_{\mathrm{C}}$. }%
\label{fig2}%
\end{figure}

Here it has to be integrated once for each of the regions $D_{i}$ (see
(\ref{D})) with initial conditions
\begin{equation}
n_{e}=n_{\gamma}=\rho_{e}=\rho_{\gamma}=\pi_{e\parallel}=0;\quad
\mathcal{E}_{0}=\tfrac{Q}{r_{i}^{2}}.
\end{equation}
Let us recall the main results of the numerical integration. The system
undergoes plasma oscillations:

\begin{enumerate}
\item  the electric field oscillates with lower and lower amplitude around $0$;

\item  electrons and positrons oscillates back and forth in the radial
direction with ultrarelativistic velocity;

\item  the oscillating charges are trapped in a thin shell whose radial
dimension is given by the elongation $\Delta l=\left|  l-l_{0}\right|  $ of
the oscillations, where $l_{0}$ is the radial coordinate of the centre of
oscillation and
\begin{equation}
l=\int_{0}^{t}\tfrac{\pi_{e\parallel}}{\rho_{e}}dt.
\end{equation}
Note that $\tfrac{\pi_{e\parallel}}{\rho_{e}}\equiv v$ is the radial mean
velocity of charges (we plot the elongation $\Delta l$ as a function of time
in Fig.~\ref{fig2});

\item  the lifetime $\Delta t$ of the oscillation is of the order of
$10^{2}-10^{4}\tau_{\mathrm{C}}$ (see Fig.~\ref{fig2}).

\item  in the time $\Delta t$ the system thermalizes in the sense that both
number and energy equipartition between electron--positron pairs and photon
are approached.
\end{enumerate}

\begin{figure}[th]
\epsfxsize=10cm \centerline{\epsfxsize=3.9in\epsfbox{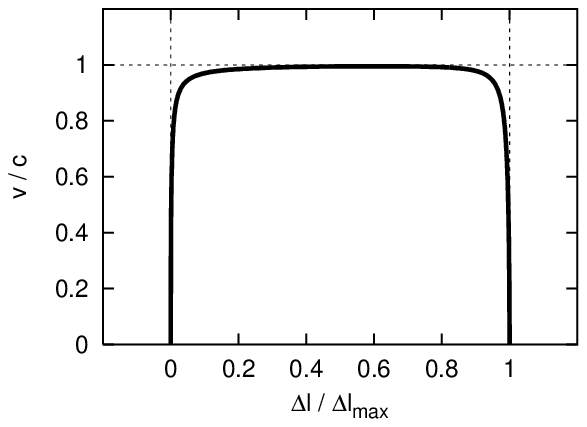}}%
\caption{Electrons mean velocity as a function of the elongation during the
first half oscillation. The plot summarize the oscillatory behaviour: as the
electrons move, the mean velocity grows up from $0$ to the speed of light and
then falls down at $0$ again. }%
\label{fig3}%
\end{figure}

In Fig.~\ref{fig3} we plot electrons mean velocity $v$ as a function of the
elongation during the first half period of oscillation, which shows precisely
the oscillatory behaviour.

\section{Conclusions}

In a paper under preparation\cite{RVX03} we are examining the conditions under
which the charge of the collapsing core is not annihilated due to vacuum
polarization as a consequence of the above plasma oscillations.

Note that $e^{+}e^{-}\rightleftarrows$ $\gamma\gamma$ scatterings is marginal
at early times ($t\ll\Delta t$) since the cross sections $\sigma_{1,2}$ are
negligible in the beginning of pair production.\cite{RVX02a} However at late
times ($t\gtrsim\Delta t$) the system is expected to relax to a plasma
configuration of thermal equilibrium.\cite{RVX02a} Thus a regime of
thermalized electrons-positrons-photons plasma begins in which the system can
be described by hydrodynamic equations. It is shown in Refs.~\refcite
{RSWX00,RVX02b} that the equations of hydrodynamic imply the expansion of the
system. In ``brief'' the system reaches the ultrarelativistic velocities
required in a realistic model for GRBs. It is worthy to remark that the time
scale of hydrodynamic evolution ($t\sim0.1s$) is, in any case, much larger
than the time scale $\Delta t$ needed for thermalization.


\begin{thebibliography}{99}
\bibitem{PRX98}G. Preparata, R. Ruffini and S.-S. Xue, \textit{A\&A}
\textbf{338}, L87 (1998).

\bibitem{HE35}W. Heisenberg and H. Euler, \textit{Zeits. Phys.} \textbf{98}
(1935) 714.

\bibitem{S51}J. Schwinger, \textit{Phys. Rev.} \textbf{82}, 664 (1951).

\bibitem{RBCFX01a}R. Ruffini, C. L. Bianco, P. Chardonnet, F. Fraschetti and
S.-S. Xue, \textit{ApJ} \textbf{555}, L107 (2001).

\bibitem{RBCFX01b}R. Ruffini, C. L. Bianco, P. Chardonnet, F. Fraschetti and
S.-S. Xue, \textit{ApJ} \textbf{555}, L113 (2001).

\bibitem{RBCFX01c}R. Ruffini, C. L. Bianco, P. Chardonnet, F. Fraschetti and
S.-S. Xue, \textit{ApJ} \textbf{555}, L117 (2001).

\bibitem{RVX03}R. Ruffini, L. Vitagliano and S.-S. Xue, in preparation.

\bibitem{CRV02}C. Cherubini, R. Ruffini and L. Vitagliano, \textit{Phys. Lett.
} \textbf{B545}, 226 (2002).

\bibitem{KESCM91}Y. Kluger, J. M. Eisenberg, B. Svetitsky, F. Cooper and E.
Mottola, \textit{Phys. Rev. Lett.} \textbf{67}, 2427 (1991).

\bibitem{KESCM92}Y. Kluger, J. M. Eisenberg, B. Svetitsky, F. Cooper and E.
Mottola, \textit{Phys. Rev.} \textbf{D45}, 4659 (1992).

\bibitem{CEKMS93}F. Cooper, J. M. Eisenberg, Y. Kluger, E. Mottola and B.
Svetitsky, \textit{Phys. Rev.} \textbf{D48}, 190 (1993).

\bibitem{RSWX00}R. Ruffini, J. D. Salmonson, J. R. Wilson and S.-S. Xue,
\textit{A\&A} \textbf{359}, 855 (2000).

\bibitem{RVX02a}R. Ruffini, L. Vitagliano and S.-S. Xue, \textit{Phys. Lett.}
\textbf{B559} (2003) 12.

\bibitem{RVX02b}R. Ruffini, L. Vitagliano and S.-S. Xue, (2003) \textit{in
these proceedings}.
\end{thebibliography}
\end{document}